# Atomic layer etching of SiO$_2$ using sequential SF$_6$ gas and Ar plasma

Jun Peng[1]*, Rakshith Venugopal[1], Robert Blick[1,2], Robert Zierold[1]*.

1  Center for Hybrid Nanostructures, University of Hamburg, Luruper Chaussee 149, 22607 Hamburg, Germany

2  Deutsches Elektronen-Synchrotron (DESY), Notkestr. 85, 22607 Hamburg, Germany

*: Corresponding authors: jpeng@physnet.uni-hamburg.de, rzierold@physnet.uni-hamburg.de

## Abstract

In the relentless pursuit of advancing semiconductor technologies, the demand for atomic layer processes has given rise to innovative processes, which have already played a significant role in the continued miniaturization features. Among these, atomic layer etching (ALE) is gaining increasing attention, offering precise control over material removal at the atomic level. Despite some thermal ALE achieved sub-nm etching controllability, the currently practical ALE processes that involve plasmas steps often suffer from high etch rates due to the scarcity of highly synergistic ALE half-reactions. To overcome this limitation, we developed an ALE process of silicon dioxide (SiO$_2$) on a silicon wafer using sequential pure sulfur hexafluoride (SF$_6$) gas exposure and argon (Ar) plasma etching near room temperature, achieving a stable and consistent etching rate of approximately 1.4 Å/cycle. In this process, neither of the two half-cycle reactions alone produces etching effects, and etching only occurs when the two are repeated in sequence, which means a 100% synergy. The identification of temperature and plasma power windows further substantiates the high synergy of our ALE process. Moreover, detailed morphology characterization over multiple cycles reveals a directional etching effect. This study provides a reliable, reproducible, and highly controllable ALE process for SiO$_2$ etching, which is promising for nanofabrication processes.

## Introduction

Atomic layer processing technologies, including atomic layer deposition (ALD) and ALE, have emerged as key techniques in the semiconductor industry,[1-4] and have further been instrumental in fabricating recent generations of nanoelectronics such as quantum devices.[5-9] They offer true atomic-level control on the processing thickness, pushing the limits of feature size, three-dimensional scaling, and overall device performance, while also being compatible with wafer-scale fabrication and high-volume manufacturing.[10-12] The concept of ALE first appeared in a 1988 patent, which described the removal of single atomic layers of crystalline diamond by alternating exposure to nitrogen dioxide and bombardment with inert gas ions in a plasma.[13] However, due to the limited demand for such precise etching techniques at the time, it did not raise much attention. As Moore's Law has reached its limits in the past decade, ALE has gained more attention and has even been used in logic devices at the 10 nm technology node.[11] And in 2015, Lee and George[14] reported the first isotropic etching of $Al_2O_3$ using a thermal ALE process, inspiring further exploration of ALE as a unique dry etch technique. However, plasma-related ALEs, which is more practical and widely used, still suffer from excessively high etching rates due to non-ideal synergy between half-reactions, where other etching reactions such as physical sputtering occur. ALE synergy, $S$, can be used to check the purity of the ALE process and is quantified as a percentage relative to the total amount of material etched per cycle (EPC), $S = \frac{\text{EPC} - (\alpha + \beta)}{\text{EPC}} \times 100\%$, where is the values of "$\alpha$" and "$\beta$" are undesirable contributions from the individual surface modification steps and the removal step, respectively.[10, 15]

$SiO_2$, a critical material in the semiconductor field, has been extensively explored using various ALE strategies.[16] These include thermal ALEs with trimethylaluminum as a precursor,[17-18] plasma-related ALEs using fluorocarbons to modify the passivated surface,[19-20] and ALE utilizing pure infrared thermal effects to etch the modified surface.[21] While thermal ALE obtained an EPC below 1 Å/cycle, its isotropic etching characteristics limit its applicability in an environment where directional etching remains the primary demand. Although plasma-involved ALE offers good directional etching results, challenges such as high etching rates and weak synergy persist in $SiO_2$ ALE. Here, we demonstrate an ALE strategy for $SiO_2$, whose ALE temperature window is near room temperature, using sequential $SF_6$ gas and Ar plasma, and achieving a constant EPC of approximately 1.4 Å/cycle. The 100% synergy of this strategy was confirmed through a systematic study involving plasma power, dose, and substrate temperature. Additionally, this ALE process was used to etch pillars and holes, exhibiting excellent directional etching effects. This work provides a competitive, reliable, and controllable directional etching option for high-precision micro-nanofabrication.

**Experimental Section**

The samples used were cut from a 4-inch $SiO_2$ (300 nm)/Si wafers (SIEGERT WAFER GmbH). Each sample was cut into approximately 1×1 cm. The samples were cleaned in acetone, isopropanol and deionized water. The etching was done in a commercial reactive ion etching (RIE) System (SenTech SI 500). The samples were inserted into the chamber under vacuum. The standard process of the ALE process is a cyclic process performed at a constant 23 °C with a continuous flow of 100 sccm of Ar at a working pressure of 1 Pa. $SF_6$ is injected into the reaction chamber at a rate of 20 sccm for 5 seconds alongside 100 sccm Ar, then the injection of $SF_6$ is stopped. The excess $SF_6$ is purged away by the Ar gas, for 30 seconds. This is termed the Purge Process. After which the plasma is activated by applying an ICP power of 100 W for 60 seconds. After turning off the ICP power, another Purge Process occurs for 30 seconds. This constitutes one ALE process. It was found that the etch rate is ~1.4 Å/cyc. The thickness of the samples was measured with an *ex-situ* ellipsometer (SenTech). The thickness is determined by using a Cauchy Model for $SiO_2$. The film thickness was measured before and after etching. For the pillars, positive e-beam resist ARP661.09 was used. The pattern was defined by an electron beam lithography system (Raith). The resulting sample was then subjected to a deep ion etching in the RIE system, with a continuous flow of 60 sccm $SF_6$, a plasma was generated with an ICP power of 300W and an radio frequency (RF) bias of 60 W. The etching was carried out for 70 seconds. Then, the samples were cleaned with an Ar plasma of 300 W of ICP power for 120 seconds in the same RIE system. Scanning Electron Microscope images were taken from SEM (Zeiss Crossbeam 550), and the roughness was measured by Atomic Force Microscope (AFM, Dimension).

## Results and discussion

*Process of the proposed ALE*

According to the consensus of the ALE community, an ALE process decomposes the whole etching process into two or more individually controlled, self-limiting, surface reaction steps that remove material only when run in sequence.[11] Our ALE process is designed based on this idea, which consists of four steps per cycle: (i) surface modification step, (ii) purge step, (iii) removal step, and (iv) purge step as shown in Figure 1a. In the modification step (Figure 1a(i)), the $SF_6$ molecules are introduced into the reactor and adsorbed on the exposed substrate surface in a self-limiting manner. After a purge step (Figure 1a(ii)), the Ar plasma is activated (Figure 1a(iii)), generating Ar+ ions and free electrons. The plasma products react with the layer of $SF_6$ adsorbed on the surface, resulting in active substances, e.g., $SF_5^+$, $SF_4^{2+}$, and F-radicals.[22] These substances, especially F-radicals, are highly reactive and can react with $SiO_2$ to generate volatile byproducts such as $SiF_4$.[23] Because $SF_6$ is adsorbed on the substrate surface in a self-limiting manner, the quantity of $SF_6$ and the movement and range of action of the active substances are limited. This constraint ensures that only a single layer of the surface is etched after another purge step (Figure 1a(iv)). Note that the single "layer" etched in ALE, similar to ALD, refers to a fixed thickness etched in each cycle, typically in the sub-nm range. If the material being etched is crystalline, this thickness is usually equal to or slightly less than a crystalline monolayer thickness. In addition, in practical applications, the etched material is often amorphous, removing the monolayer concept not strictly applicable. In ALE, the thickness etched in each cycle is defined as EPC. Figure 1b shows the etch thickness for different cycle numbers. The EPC obtained by linear fitting is 1.4 Å/cycle with $R^2 \approx 0.999$. The characterization was conducted using the standard loop recipe for this work: a pulse of SF6 gas molecules (20 sccm, 5 s) for the modification step, followed by 30 s purge; then a pulse of Inductively Coupled Plasma (ICP) (100 W, 60 s) for the removal step, with another purge for 30 s. Throughout the process, Ar served as a carrier gas with a flow rate of 100 sccm. The reactor was maintained at room temperature and 1 Pa pressure. Figure 1c and Table S1 compare the EPC between this work and previous explorations over the past decade,[19, 21, 24-32] demonstrating the superiority of our "sequential $SF_6$ gas and Ar plasma" ALE strategy in accuracy. Our approach has achieved an EPC on par with thermal ALE, showcasing its potential in further research and practical applications.

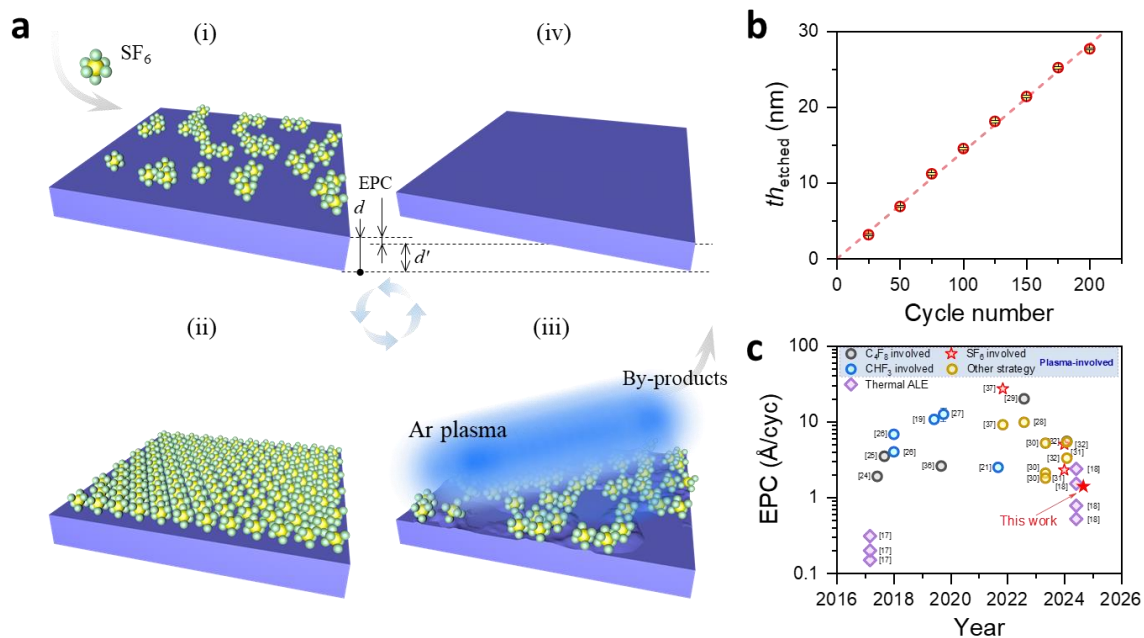

**Figure 1. ALE process.** (a) Schematic diagram of the ALE strategy using sequential $SF_6$ gas and pulsed Ar plasma. The process consists of two half-reactions, with the total cycle divided into four steps: (i) In the surface modification step, a pulse of $SF_6$ molecules is adsorbed on the exposed substrate surface in a self-limiting manner. (ii) A purge step follows to remove excess molecules, leaving a thin $SF_6$ layer on the exposed silica surface. (iii) In the removal step, a pulse of Ar plasma is applied, breaking the bonds and removing the surface atomic layer. (iv) Subsequently, the reaction chamber is purged again, leaving a new, fresh $SiO_2$ surface. Each cycle removes a thin layer of silicon dioxide from the surface, termed Etch Per Cycle (EPC). (b) EPC is defined 1.4 Å/cycle from the etched thickness, $th_{etched}$, over cycle numbers by linear fitting with $R^2 \approx 0.999$. (c) Comparison of the EPC from this work with the state-of-the-art plasma-involved $SiO_2$ ALE in the past decade. The numbers next to the data points represent the corresponding reference number. The precision of our plasma-involved process is comparable to some thermal ALE processes.

*ALE synergy characterization*

In ALE, the modification and removal steps are separated by two purge steps in the loop setup as displayed in Figure 2a. Ideally, no step in ALE will produce etching effect alone. In order to better understand the etching mechanism of this ALE process, it is worthwhile to conduct a more careful and systematic study of the control variables of the surface modification step and the removal step, and characterize their ALE synergy effect S simultaneously. First, the modification step is removed from the standard process, and the effect of ICP power (removal step) on etching is studied separately. Such a process can be seen as pure $Ar^+$ sputtering process. From the results shown in Figure 2b, it can be seen that the EPC corresponding to the four ICP powers fluctuates around zero. This fluctuation is caused by the error of the ellipsometer measurement. Therefore, within the tested ICP power range (50~100 W), $\beta$ equals zero. Next, the etching step is removed from the cycle and the $SF_6$ dose (modification step) is studied. Obviously, $SF_6$ alone does not produce any etching (Figure 2c), which means $\alpha$ equals zero. It can be calculated that the ALE synergy of the standard process in this work is $S = 100\%$. Such a result is much better than the synergy (~80%) of the ALE using alternating fluorocarbon plasma and argon ions that has been used in the production of logic devices at the 10 nm technology node.[15] ALE has an important characteristic, self-limitation, which can be verified by controlling the dose of the precursor. As shown in Figure 2d, gradually increasing the dose of $SF_6$ results in a gradual increase in EPC, which reaches saturation at around 25 sccm·s. Beyond this point, the EPC no longer changes significantly with further increases in the dose. Note that lower doses cannot be set by the instrument used.

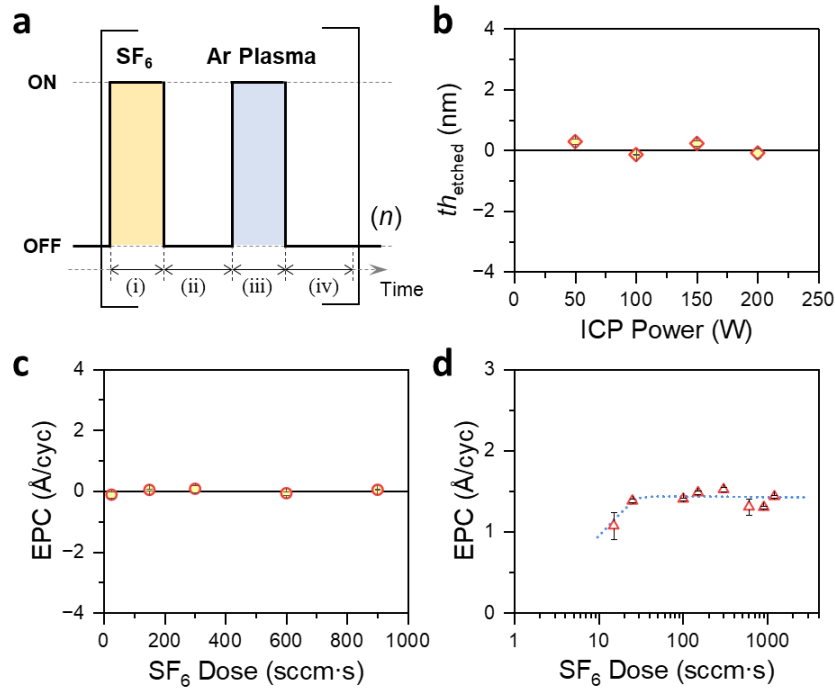

**Figure 2. The process setup and synergy verification.** (a) Process setup. By controlling the variables, the synergistic effect can be verified to show how "pure" our ALE process is. Steps (i) to (iv) correspond to (i) to (iv) in Figure 1a. (b) Process tests without (i) $SF_6$ modification step. (c) Process tests without (iii) Ar plasma removal step. These two tests were performed based on the standard process, changing only the corresponding parameter amounts shown on the x-axis, namely ICP power and $SF_6$ dose, respectively. (d) Effect of $SF_6$ dose and ICP power on etching effect. This shows that the adsorption of $SF_6$ on the substrate surface is self-limiting.

*ALE windows exploration*

Based on the above data, it can be confirmed that the "sequential $SF_6$ gas and Ar plasma" ALE process demonstrates pure ALE with 100% synergy. As ALD and ALE share many similarities, an ideal ALE should also exhibit an ALE window like that ALD has an ALD window. Figure 3a shows the effect of the wafer holder temperature on the etching rate. From room temperature to about 40 °C, the EPC remains stable, indicating the ALE window location. While beyond this range until 160 °C, the EPC gradually decreases. According to the adsorption principle, this decrease in EPC may be due to $SF_6$ molecules at higher thermal energy, leading to easier desorption from the surface and thus reducing effective etching. In addition to the temperature window, an ICP power window is also identified. As ICP power increases, the ionization rate of the gas increases, resulting in higher ion density, higher electron temperature, and more energetic plasma. When the energy of incident active particles is sufficient to remove the $SF_6$-modified surface but not the underlying $SiO_2$, it falls within the ALE power window, such as the 50-100 W range in Figure 3b. Below this window, the energy of incident particles is insufficient to remove all the modified surface, causing a decreasing EPC. However, contrary to previous reports on plasma-related ALE of $SiO_2$, our EPC decreases as the ICP power increases beyond the window. Typically, the EPC rises due to the onset of physical sputtering.[33] This difference may be explained by the increased concentration of various particles at higher ICP power, diluting the concentration of dissociated F-radicals and reducing EPC. Additionally, higher energy active particles and elevated temperatures may cause more intense collisions, leading to elastic scattering and desorption of $SF_6$ from the surface, thereby diminishing effective etching. Note that the tested ICP power has not yet reached the threshold of physical sputtering in this condition as shown in Figure 2b.

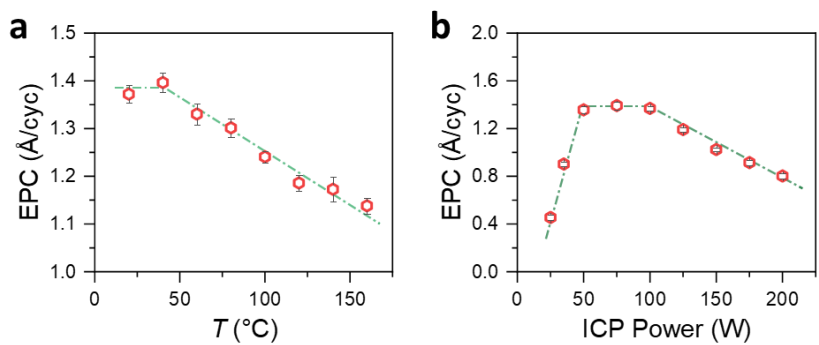

**Figure 3. Characteristics of the ALE windows.** (a) Temperature window. (b) Plasma power window. The standard process parameters are controlled as a continuous flow of 100 sccm of Ar, 5 sccm of $SF_6$ for 5 seconds, and a plasma time of 60 seconds. Only the corresponding quantities are changed during the study.

*Directional etching*

Unlike the unparalleled conformality of ALD, ALE can be categorized into directional etching and isotropic etching, each required for different applications. Therefore, verifying the type of etching in the ALE process is crucial. A substrate with pillars with 600 nm in diameter is used for testing. Figure 4a and 4b show optical microscope images and atomic force microscope (AFM) images of the sample at different stages. The color variations in the optical images result from changes in the overall film thickness due to etching. The AFM image and detailed interface profile analysis (Figure 4c) reveal that the three-dimensional surface morphology of the sample does not change significantly as the etching proceeds. This observation indicates that our ALE process is directional and does not etch the side facets of the pillars (Figure 4d). In isotropic etching, besides the pillar height remaining unchanged, the pillar diameter should uniformly reduce (Figure S2). For this sample, the entire sample was uniformly etched by 62 nm (Figure 4e). If the process were isotropic, the diameter of the pillar should have reduced by about 27%, but the actual diameter remained constant. The directional etching in our ALE process is likely the result of the bias voltage generated by factors such as the self-bias effect and potential capacitive coupling from the ICP power. Although the bias voltage was not deliberately activated during the experiment, it is still observed when the Ar plasma pulse is initiated (Figure S3). This bias voltage creates an electric field in the plasma sheath near the substrate surface, accelerating ions toward the substrate. Therefore, the bombardment of charged particles in the plasma, including $Ar^+$ ions, on the substrate is directional, enhancing the perpendicular component of their interaction with the substrate plane. For different incident angles, the energy of the particles varies significantly. The higher the incident angle, the smaller the particle energy, and the less likely it is to etch the micro-nano pillars and hole sidewalls.[34-35] At low pressure, particles also have a larger mean free path. This directional movement and angle-dependent energy ensure that etching primarily affects the vertical dimensions of the sample rather than the horizontal. Furthermore, this directional etching also shows high repeatability in hole samples (Figure S4). Additionally, after ALE etching, the sample's roughness (Ra) remained around 0.7 nm, confirming the gentle etching nature of the etching process.

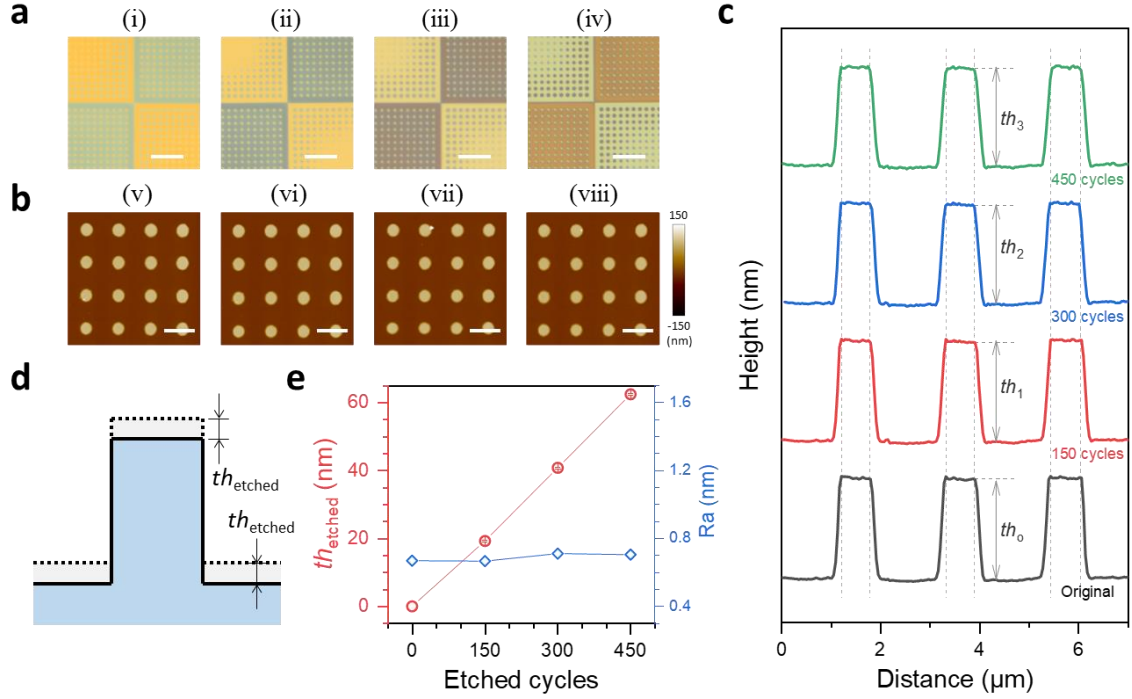

**Figure 4. Directional ALE etching.** Pillars are prepared for this characterization. (a) Optical microscope images and (b) the corresponding AFM images of the original pillar sample (i, v), the same position after 150 ALE cycles (ii, vi), the same position after 300 ALE cycles (iii, vii), and the same position after 450 ALE cycles (iv, viii). The color difference is due to the different thicknesses of $SiO_2$ after etching. The scalebar for (a), (b) are 10 μm and 2 μm. (c)The analysis of the same three consecutive pillars in their original state and after 150, 300, and 450 ALE cycles. The corresponding pillar heights $th_0$, $th_1$, $th_2$, and $th_3$ are 91.4 ± 1.17, 91.3 ± 0.86, 90.7 ± 0.91, 89.6 ± 1.00 nm, which keeps the same during the etching. (d) Sketch of the directional etching result. (e) Corresponding etched thickness $th_{etched}$ and roughness $Ra$ during the test. The stable and low $Ra$ suggests a damage-free surface.

## Conclusions

In summary, we have developed an ALE process for $SiO_2$ using sequential pure $SF_6$ gas exposure and Ar plasma etching near room temperature, achieving a stable and consistent etching rate of approximately 1.4 Å/cycle. No etching effect was observed in the individual half-reaction cycles, and we identified both an ALE temperature window near room temperature and an ALE plasma power window, indicating 100% synergy between the half-reactions. Additionally, the diameter of the micro-nano pillars and holes on the plane remained unchanged during the etching process, confirming that the process is directional. Our etching process is performed using commercial RIE equipment with commonly used gases in the semiconductor field, offering good scalability and versatility. While we used $SiO_2$ as an example to demonstrate the precise etching capability of this method, the process described here can potentially be extended to other materials that are reactive to an etching gas's plasma but unreactive to its pure gas.

## Acknowledgements

This work was funded by the Deutsche Forschungsgemeinschaft (DFG, German Research Foundation) – Projektnummer 192346071 – SFB 986 "Tailor-Made Multi-Scale Materials Systems".

## Data availability

The data supporting the findings of this work are available within the article and its Supplementary Information files. All other relevant data supporting the findings of this study are available from the corresponding author on request.

## Author contributions

J.P. proposed the concepts and designed the experiments; V.R. and J.P. performed the experiments and analyzation; R.Z. supervised the study; R.B. provided infrastructure to conduct the experiments; J.P., V.R. and R.Z. co-wrote the manuscript. All authors discussed the experimental and theoretical results and commented on the manuscript. All authors have approved the final version of the manuscript.

## Competing interests

The authors declare no competing interests.

**Table of Contents**

**Atomic layer etching of SiO$_2$ via sequential SF$_6$ gas and Ar plasma**

*Jun Peng\*, Rakshith Venugopal, Robert Blick, Robert Zierold\**

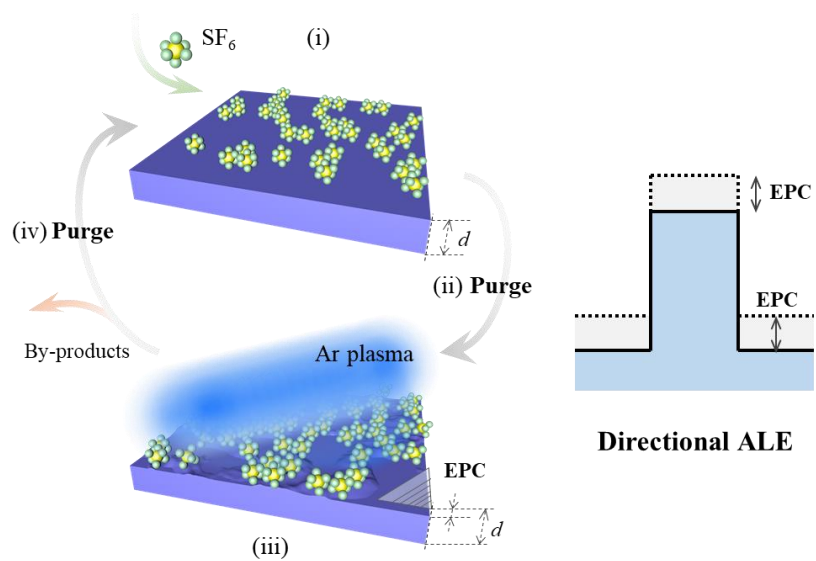

This work demonstrates an atomic layer etching process on silicon dioxide using sequential SF$_6$ gas and Ar plasma at room temperature, achieving a stable and consistent etching rate of approximately 1.4 Å/cycle. The systematic study confirms its 100% synergy and reveals its directional etching characteristics.

Supporting Information

# Atomic layer etching of SiO$_2$ using sequential SF$_6$ gas and Ar plasma

Jun Peng[1]*, Rakshith Venugopal[1], Robert Blick[1,2], Robert Zierold[1]*.

1 Center for Hybrid Nanostructures, University of Hamburg, Luruper Chaussee 149, 22607 Hamburg, Germany

2 Deutsches Elektronen-Synchrotron (DESY), Notkestr. 85, 22607 Hamburg, Germany

*: Corresponding authors: jpeng@physnet.uni-hamburg.de, rzierold@physnet.uni-hamburg.de

This supporting information file includes:

    Figure S1 to Figure S4

    Table S1

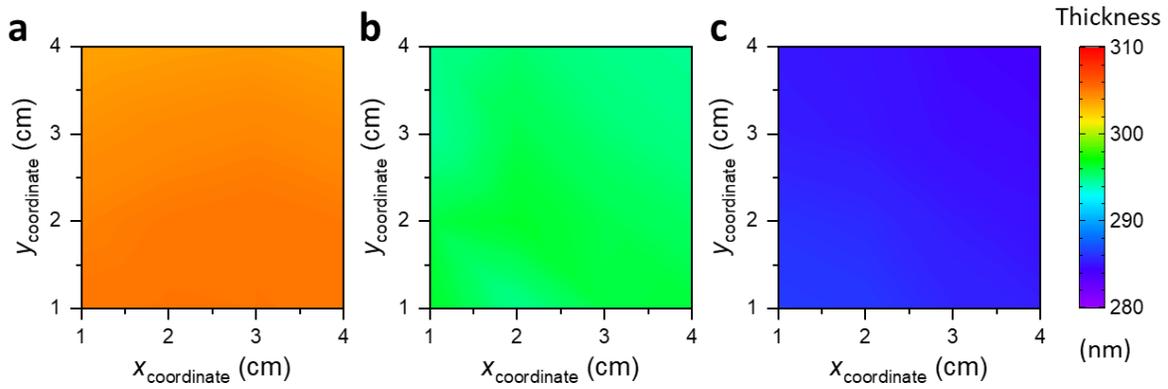

**Figure S1.** Uniformity test on the same SiO$_2$/Si wafer with a 4.5 cm × 4.5 cm size. (a) The thickness of the SiO$_2$ layer of the original sample is 304.7 ± 0.54 nm. (b) The thickness of the SiO$_2$ layer after 75 cycles is 295.5 ± 0.67 nm, and 9.2 ± 0.51 nm is etched. (c) The thickness after another 75 cycles, in total of 150 cycles, is 285.2 ± 0.61 nm, and 19.5 ± 0.50 nm is etched. The test was performed at a 4 cm × 4 cm area with a spacing of 1 cm. Each measurement was made at the same location. The standard deviations of the etched thickness after 75 cycles and 150 cycles are all around 0.5 nm, indicating good intra-wafer uniformity of the etching.

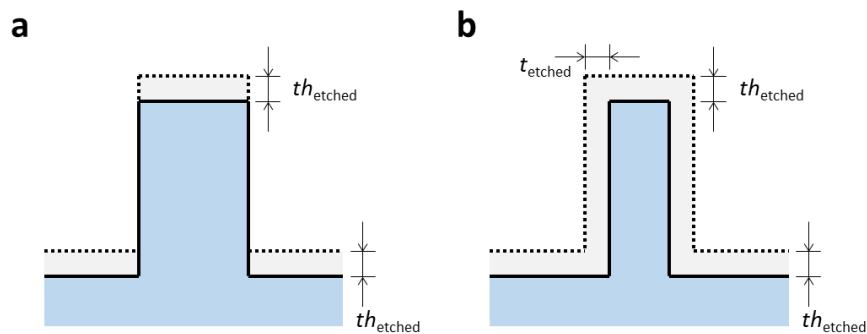

**Figure S2.** Comparison of directional etching and isotropic etching. (a) Directional etching will only etch in the vertical direction, not in the horizontal direction. (b) However, in isotropic etching, the cylinder will be uniformly etched in both the vertical and horizontal directions, resulting in a reduction in diameter.

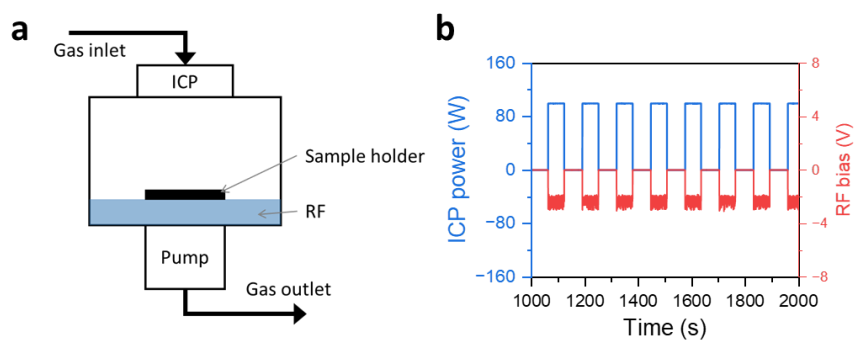

**Figure S3.** Indirect bias voltage. (a) The sketch of the reactor. (b) Part of the instrument log showing the detected ICP power and RF power during the experiment. During the experiment, only the ICP on the top of the sample was activated in the form of pulses, and the RF power was set to zero. Every time the ICP was activated, a passively generated RF bias was recorded. This bias will make the plasma directional.

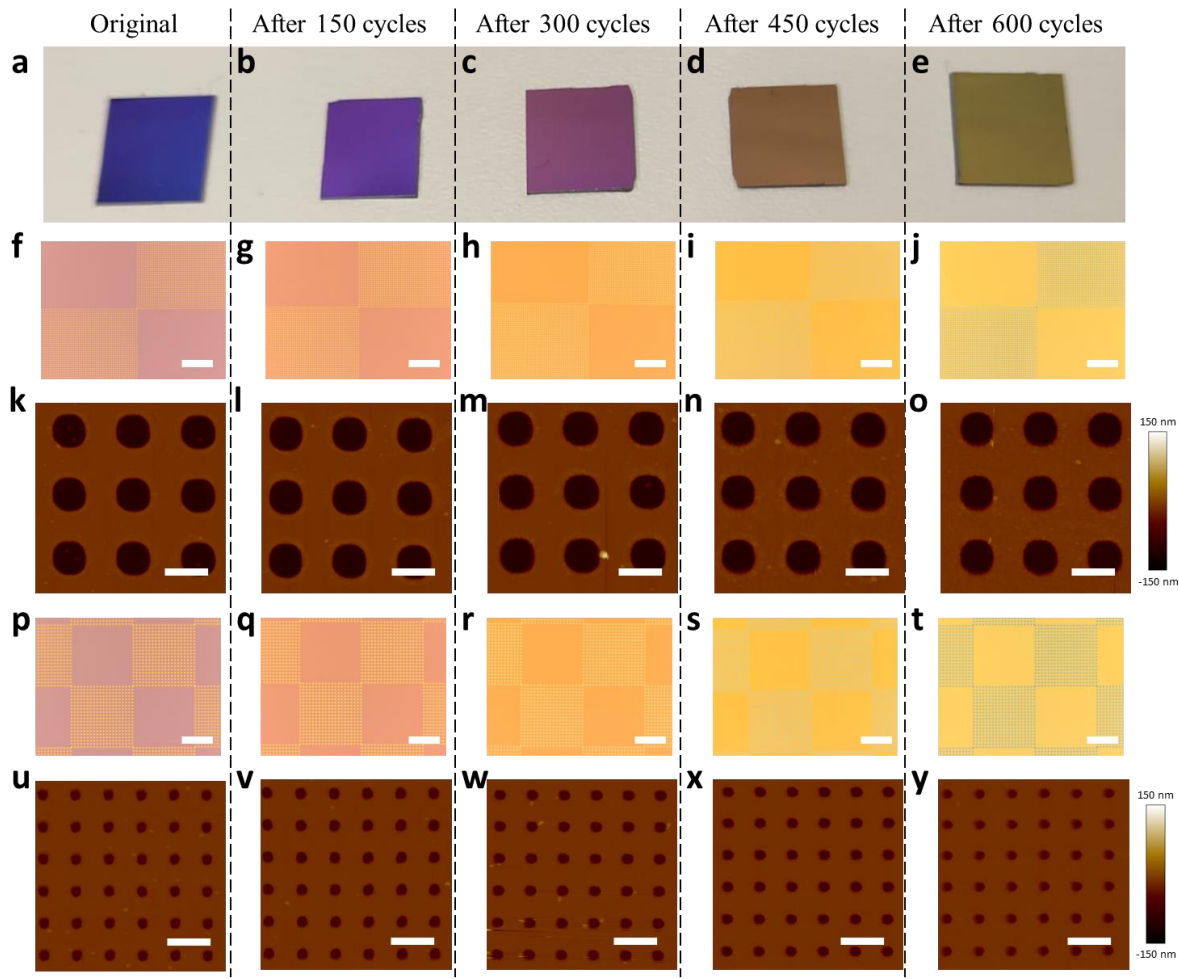

**Figure S4.** Holes sample for the etching effect test. (a~e) Photo of the etching reference samples. The size of the five reference samples is about 1 cm × 1 cm. Samples (b)~(e) were placed in the reaction chamber at the first batch and taken out after the corresponding etching batch in turns. (f~j) Optical images of the same sample, with a hole diameter of about 1.2 μm, at the same position after different number of etching cycles. (k~o) AFM images of the same sample after different number of etching cycles corresponding to the sample in (f~j). (p~t) Optical images of the same sample, with a hole diameter of about 0.6 μm, at the same position after different number of etching cycles. (u~y) AFM images of the same sample after different number of etching cycles corresponding to the sample in (p~t). Etching reduces the thickness of the $SiO_2$ film, causing the color of the film to change, but the diameter of the pores does not change. The scalebar for (f~j), (k~o), (p~t), and (u~yt) are 20 μm, 2 μm, 20 μm, and 2 μm.

**Table S1.** EPC comparisons of reported $SiO_2$ etching using ALE manner.

| Time | Reactant A | Reactant B | Further reactants | EPC (Å/cyc) | ref |
|---|---|---|---|---|---|
| 2017-02 | Trimethylaluminum ($Al(CH_3)_3$) | HF | -- | 0.027 (0.1 Torr) 0.15 (0.5 Torr) 0.2 (1 Torr) 0.31 (4 Torr) | 17 |
| 2017-05 | $C_4F_8$ plasma | Ar plasma | -- | 1.9 | 24 |
| 2017-08 | $C_4F_8$/Ar plasma | Ar plasma | -- | 3 ~ 4 | 25 |
| 2017-12 | $CHF_3$ plasma | $O_2$ or Ar plasma | -- | 6.8 ($O_2$ plasma) 4.0 (Ar plasma) | 26 |
| 2019-05 | Ar plasma | $CHF_3$ | -- | 10.7 | 19 |
| 2019-08 | $C_4F_8$/Ar plasma | Ar plasma | -- | 2.6 | 36 |
| 2019-09 | $CHF_3$ | Ar Plasma | -- | 10 ~ 15 | 27 |
| 2021-08 | $CHF_3/O_2$ plasma | infrared annealing |  | 2.5 | 21 |
| 2021-10 | HF | $NH_3$ | infrared annealing | 9.09 | 37 |
| 2021-10 | $H_2$, $SF_6$ plasma | $NH_3$ | infrared annealing | 27.0 | 37 |
| 2022-07 | $CF_3I$ plasma | $O_2$ plasma | -- | 9.8 | 28 |
| 2022-07 | $C_4F_8$ | Ar Plasma | -- | 20 | 29 |
| 2023-04 | Heptafluoropropyl methyl ether (HFE-347mcc3) | Ar plasma | -- | 2.1 | 30 |
| 2023-04 | Heptafluoroispropyl methyl ether (HFE-347mmy) | Ar plasma | -- | 1.8 | 30 |
| 2023-04 | Perfluoro propyl carbinol (PPC) | Ar plasma | -- | 5.2 | 30 |
| 2023-12 | $SF_6$ plasma | Ar plasma | -- | 2.3 (without bias) 5 (with bias) | 31 |
| 2024-01 | $C_4F_8$ | Ar plasma | -- | 5.5 | 32 |
| 2024-01 | perfluoroisopropyl vinyl ether (PIPVE) | Ar plasma | -- | 3.3 | 32 |
| 2024-01 | perfluoropropyl vinyl ether (PPVE) | Ar plasma | -- | 5.4 | 32 |
| 2024-05 | Trimethylaluminum ($Al(CH_3)_3$) | $Ar/H_2/SF_6$ plasma | -- | 0.52 ($SiO_2$) 0.78 (ALD $SiO_2$) 1.52 (PECVD) 2.38 (Sputtered) | 18 |
| ★ | $SF_6$ | Ar plasma | -- | 1.4 | This work |